\documentclass[sigconf]{acmart}

\AtBeginDocument{%
  \providecommand\BibTeX{{%
    \normalfont B\kern-0.5em{\scshape i\kern-0.25em b}\kern-0.8em\TeX}}}

\copyrightyear{2022} 
\acmYear{2022} 
\setcopyright{rightsretained} 
\acmConference[CUI 2022]{4th Conference on Conversational User
Interfaces}{July 26--28, 2022}{Glasgow, United Kingdom}
\acmBooktitle{4th Conference on Conversational User Interfaces (CUI
2022), July 26--28, 2022, Glasgow, United
Kingdom}
\acmDOI{10.1145/3543829.3544511}
\acmISBN{978-1-4503-9739-1/22/07}

\begin{document}

\title[Bilingual by default]{Bilingual by default: Voice Assistants and the role of code-switching in creating a bilingual user experience}

\author{Helin Cihan}
\affiliation{
  \institution{University College Dublin}
  \city{Dublin}
  \country{Ireland}}
  \email{esna.ogutcucihan@ucdconnect.ie}

\author{Yunhan Wu}
\affiliation{
 \institution{University College Dublin}
 \city{Dublin}
 \country{Ireland} }
 \email{yunhan.wu@ucdconnect.ie}

\author{Paola Peña}
\affiliation{
 \institution{University College Dublin}
 \city{Dublin}
 \country{Ireland}}
 \email{paola.pena@ucdconnect.ie}

\author{Justin Edwards}
\affiliation{
 \institution{University College Dublin}
 \city{Dublin}
 \country{Ireland}}
 \email{justin.edwards@ucdconnect.ie}

\author{Benjamin R. Cowan}
\affiliation{
 \institution{University College Dublin}
 \city{Dublin}
 \country{Ireland} }
 \email{benjamin.cowan@ucd.ie}


\begin{abstract}
  Conversational User Interfaces such as Voice Assistants are hugely popular. Yet they are designed to be monolingual by default, lacking support for, or sensitivity to, the bilingual dialogue experience. In this provocation paper, we highlight the language production challenges faced in VA interaction for bilingual users. We argue that, by facilitating phenomena seen in bilingual interaction, such as code-switching, we can foster a more inclusive and improved user experience for bilingual users. We also explore ways that this might be achieved, through the support of multiple language recognition as well as being sensitive to the preferences of code-switching in speech output.
\end{abstract}

\ccsdesc[500]{Human-centered computing~Natural language interfaces}
\ccsdesc[500]{Human-centered computing~Accessibility design and evaluation methods}

\keywords{speech interface, voice user interface, intelligent personal assistants, non-native speakers, code-switching}

\maketitle
\section{Introduction and Motivation}
Today, most of the world’s population can speak more than one language \cite{grosjean_bilingual_2012}.  As conversational user interfaces (CUIs) proliferate rapidly, there is an opportunity for voice assistants to become sensitive to this fact, supporting bilingual interaction. CUIs such as voice assistants (VAs) are regularly used in one on one interactions as well as in situations where there are multiple users conversing with the VA as well as each other (e.g. \cite{porcheron2018voice}). Current VAs are designed to be monolingual by default, interacting with users based on the language they use to converse with the system or based on the language selected by the user. Indeed many VA users tend to engage in their second language, or risk being excluded from VA based functionality as functional coverage is not uniform \cite{wu_see_2020}. Yet bilinguals do not tend to engage in only one language when conversing with each other. For instance, both low and high proficient bilinguals commonly change between languages while speaking, sometimes within one utterance \cite{sankoff_formal_1981, isurin_multidisciplinary_2009}. This mixing of languages, or code-switching, occurs when speakers replace words, phrases or sentences in one language with another during communication \cite{li_spoken_1996, grosjean_listening_2018}, and is one of the most prominent phenomena that arise in bilingual conversation \cite{kootstra_code-switching_2012,DBLP:journals/corr/abs-1904-00784}. This provocation paper argues that, when interacting with bilinguals or in bilingual groups, rather than being monolingual by default, our voice assistants need to move from being monolingual to become multilingual by default. VAs should not only become sensitive to the patterns of language use we see in multilingual communities but indeed adopt these same patterns when conversing with bilingual and multilingual users and groups. Only then will VAs be able to support the interaction patterns in multilingual interaction. Our paper specifically highlights the challenges faced when talking in a second language, focusing on the concept of code-switching as a way for VAs to start facilitating bilingual dialogue experiences. 

\section{The challenge of interacting in a second language }
Bilingual speakers are skilled linguistic multitaskers who routinely face cognitive challenges while speaking. For instance, when bilingual people communicate, both language systems become activated \cite{kroll_automaticity_2005,kootstra_code-switching_2012, costa_bilingualism_2008} and this simultaneous activation obliges the bilingual speaker to keep two languages separate during speech by inhibiting their non-target language \cite{abutalebi_bilingual_2007, guo_local_2011,bialystok_bilingualism_2012}. Although this continuous control is known to induce beneficial effects on certain non-verbal cognitive skills (e.g. \cite{bialystok_bilingualism_2004, martin-rhee_development_2008, costa_bilingual_2009, hernandez_impact_2010}) verbal skills of bilinguals in each of their languages tend to be weaker than those of monolinguals in either language \cite{bialystok_bilingualism_2012}, with both bilingual children and adults possessing a smaller receptive vocabulary size in each of their languages in comparison to monolingual peers in a given language \cite{bialystok_receptive_2010, bialystok_receptive_2012}. Even among highly proficient bilinguals this language co-activation may result in less efficient non-dominant (L2) language production compared to dominant language (L1). Lexical access disadvantages in bilinguals during language production compared to monolinguals may also present themselves even in bilingual speakers’ L1 \cite{ivanova_does_2008, branzi_cross-linguisticbilingual_2018}. In sum, retrieval from one’s lexicon for production is more effortful for bilinguals compared to monolinguals.
 
\section{The language use problem in L2 CUI interaction}
Interacting with speech agents such as Google Assistant, Amazon Alexa, and Apple’s Siri has become ubiquitous. Currently not all languages are supported across speech agents \cite{kinsella_2019}, which forces many users to use their L2 rather than their L1 to interact with speech interfaces. Recent research has investigated the challenges L2 speakers face in VA interaction \cite{pyae_19,pyae_18,wu_mental_2020,wu_see_2020}. Some of these works have focused especially on bilingual speakers with an L1 that differs greatly in phonological, lexical and syntactic systems from their L2, such as Mandarin and English \cite{wu_mental_2020,wu_see_2020}. In such cases, L2 speakers exert additional mental effort in producing accurate L2 speech during interaction \cite{watson2013effect}. For some bilingual speakers, strictly interacting with a VA in an L2 monolingual interaction significantly increases mental effort compared to an L1 monolingual interaction \cite{wu_mental_2020}, and such an environment may lead them to disengage \cite{green_language_2013}. In this way, VA interactions as they currently exist disadvantage bilingual speakers, making language use more challenging by limiting interaction to a single language in which they are sometimes not perfectly proficient. Rather than hindering, we feel that VAs should support bilingual behaviours, leveraging phenomena frequently seen in bilingual interactions. Below we focus in particular on how supporting the phenomena of code-switching could support bilingual users.

\section{Code-switching recognition can improve CUI understanding} 
Code-switching allows bilinguals to shuttle between two languages, and in some cases, this voluntary use may be less effortful than confinement to one language \cite{de_bruin_not_2019, grosjean_listening_2018}. Bilingual speakers may face particular challenges in L2 VA interaction, not because they do not know a particular word, but that they have a hard time retrieving it from their lexicon due to its infrequent use \cite{heredia_bilingual_2001},  especially when they are tired or distracted and try to compensate for their deficiency \cite{crystal_cambridge_2019, skiba_code_1997}. Supporting simultaneous recognition of multiple languages by the VA would lead to easier and faster language production for bilingual speakers who  experience retrieval difficulty or due to a low degree of proficiency, face more effortful L2 production \cite{dornyei_problem-solving_1998,heredia_bilingual_2001,segalowitz_automaticity_2005, de_bruin_not_2019}. Facilitating code-switching in this context would allow bilinguals to use the language that they felt most easily allowed them to produce specific parts of commands that they are finding hard to retrieve. Such recognition functionality would be particularly helpful for low-proficient bilinguals who use their non-dominant language to interact with VAs.

For bilingual speakers, especially those confined to using VAs only in their L2, recognition of speech that utilises code-switching presents a major opportunity to decrease the cognitive burden of language production seen in L2 VA interaction \cite{wu_mental_2020}. Recent work has already demonstrated the practical feasibility of recognising code-switched natural language in text-based chatbots using combinations of both Spanish and English as well as Hindi and English \cite{ahn_what_2020, parekh_understanding_2020}. These studies demonstrated high user satisfaction and conversational success when bilingual users were empowered to use code-switching while interacting with a chatbot, suggesting that this is indeed a rich vein for the improvement of CUIs more generally. 

Yet recognizing code-switched speech is no easy task of course. Bilinguals differ in their use of code-switching, some of them may be highly prone to use this switch bidirectionally, whereas others may use a unidirectional approach \cite{antoniou_inter-language_2011}. Likewise, analysis of code-switching between Spanish and English has revealed some strategies that individual bilingual speakers tend to use which may be idiosyncratic to particular language pairs \cite{bullock_should_2018}. Identifying these strategies for a given pair of languages can act as the first step toward bilingual speech recognition, as demonstrated by Ahn et al. \cite{ahn_what_2020}. Cues for code-switching strategies like syntactic structures, lexical choice, gestures, and voice intonation are sent and recognized constantly by bilingual speakers as they align their speech patterns in natural dialogue \cite{li_spoken_1996, backus2005codeswitching}. Furthermore, code-switching is easier if both languages have the same word order \cite{kootstra_code-switching_2012}. For instance, Backus \cite{backus2005codeswitching} gives the illustrative example of German spoken in Australia, where English is the dominant language and code-switching between the two languages is common. Australian German sentences have taken on English syntactic patterns, following a subject-verb-object (SVO) structure rather than the verb-second in standard German, owing to the increased ease of code-switching presented by matching syntaxes. For this reason, it may be easier for VAs to support code-switching between similar languages, with distant languages presenting a larger challenge. Even a general understanding of the strategies people use to code-switch can help with this challenge however, as work on a chatbot that can recognize code-switched language including English and Hindi - a language with a different syntactic structure than English - has demonstrated \cite{parekh_understanding_2020}. It is clear that improvements to how we understand code-switching behaviour can lead to improvements in how VAs understand code-switching. For this reason, we believe more focus on code-switching as a language behaviour, including the implementation of code-switching recognition in VAs, is necessary for accomplishing the goal of increasing the accessibility of VAs to bilingual speakers.

\section{Voice Assistants as Code Switchers}
For some, code-switching might be preferred to aid in the production of language. Yet others may also want their VAs to code-switch in their speech output, making it easier to understand VA speech output whilst making VA interactions more akin to human-human bilingual interactions. We therefore propose that VAs should not only recognise code-switching but that they should also allow users to choose whether or not their interface will code-switch when producing speech. Accessible and usable speech interfaces should optionally perform code-switching or produce utterances monolingually based on the preferences of the user while recognizing code-switched speech, in order to support users with diverse code-switching preferences. For those who want their VA to code-switch, designers must decide how the VA should coordinate that code-switching. We propose that code-switching strategies could be accounted for by observing how and when users employ coordinate their code-switching when interacting with the agent, followed by implementing linguistic alignment or entrainment \cite{pickering_toward_2004,bevnuvs2011pragmatic,levitan2016implementing} by the VA in the conversation. While some studies applied code-switching strategies in chatbots \cite{ahn_what_2020, parekh_understanding_2020}, future VAs might adapt more dynamically, aligning with users’ code-switching strategies as they occur. Just as bilingual speakers can decode simultaneous bilingual speech, we propose that speech interactions with VAs would be more natural and efficient, if they were capable of employing similar decodable skills. 
 
\section{Conclusion}
Speech-based conversational systems are becoming immensely popular whereas users are still excessively burdened by the imposed monolingualism of VA interaction. VA technologies, as well as CUIs more broadly, must consider the different ways people use language, particularly the ways bilingual speakers adapt their language production, which has largely been ignored in the development of speech interfaces to date. Doing so will make speech interfaces accessible to new audiences and improve the user experience of a significant portion of the population. We envision code-switching VAs as a key driver of a more successful human-machine dialogue in the future. Improving CUIs, and particularly speech interfaces in this way will improve the user experience of many current users, and make this already popular technology accessible to an even wider audience around the world.

\begin{acks}
This work was conducted with the financial support of the UCD
China Scholarship Council (CSC) Scheme grant No. 201908300016,
Science Foundation Ireland ADAPT Centre under Grant No.
13/RC/2106 and the Science Foundation Ireland Centre for Research
Training in Digitally-Enhanced Reality (D-REAL) under Grant No.
18/CRT/6224.
\end{acks}

\bibliographystyle{ACM-Reference-Format}
\bibliography{Code-switching}

\end{document}